\documentclass[
aps,%
12pt,%
final,%
notitlepage,%
oneside,%
onecolumn,%
nobibnotes,%
nofootinbib,%
superscriptaddress,%
noshowpacs,%
centertags]%
{revtex4}
\usepackage{float}
\usepackage{journals}
\usepackage{graphicx}	
\usepackage{amsmath}	
\usepackage{amssymb}	
\usepackage{bm}		
\usepackage{pdflscape}	
\usepackage{subfigure}
\usepackage{longtable}
\usepackage{supertabular}
\usepackage{array}
\usepackage{changepage}
\usepackage{pdflscape}	
\usepackage[english]{babel}
\usepackage[T2A]{fontenc}            
\usepackage[utf8x]{inputenc}

\begin{document}

\noindent {\it arXiv.org version}
\bigskip\bigskip  \hrule\smallskip\hrule
\vspace{35mm}

\title{Star Formation in the Low Surface Brightness Galaxies: a Comparison with the Outer Regions of Normal Galaxies}

\author{\bf \copyright $\:$ 2026
\quad \firstname{A.V.}~\surname{Zasov}}
\email{zasov@sai.msu.ru}
\affiliation{Faculty of Physics, Lomonosov Moscow State University, Leninskie Gory, Moscow 119991, Russian Federation}
\affiliation{Sternberg Astronomical Institute, Moscow State University, Universitetsky pr., 13, Moscow 119234, Russia}

\author{\bf \firstname{N.A.}~\surname{Zaitseva}}
\affiliation{Sternberg Astronomical Institute, Moscow State University, Universitetsky pr., 13, Moscow 119234, Russia}

\author{\bf \firstname{A.S.}~\surname{Saburova}}
\affiliation{Sternberg Astronomical Institute, Moscow State University, Universitetsky pr., 13, Moscow 119234, Russia}

\begin{abstract}

\textbf{Abstract}
Low surface brightness galaxies (LSB-galaxies) and the outer regions of disks of ``normal'' surface brightness galaxies (HSB-galaxies) are compared in terms of the radial distributions of star formation rates and efficiencies, as well as the Toomre stability parameter $Q_{T}$, using data from several samples of galaxies with high $HI$ content. It is shown that the dependences of the star formation rate $SFR$ on the $HI$ surface density and the star formation efficiency $SFE$ on the disk optical brightness $\mu_{g}$ for LSB- and HSB-galaxies have a similar character, suggesting similarity in the conditions that determine the star formation rate in low gas density regions. The considered dependences persist out to distances corresponding to brightnesses of 26–27 mag/arcsec$^2$, where the expected volume density of the equilibrium gas layer should be very low. It is suggested that the disk density at large distances from the center may be significantly underestimated. 

{\bf keywords:} galaxies: spiral, galaxies: LSB, galaxies: star formation.

\end{abstract}

\maketitle

\section{Introduction}

Disk galaxies with low surface brightness (LSB-galaxies) are usually defined as galaxies whose stellar disk brightness  is around or below the brightness of the dark night sky (22–23 mag/arcsec$^2$ in the B-band) even in the central regions. However, a total luminosity range of LSB-galaxies is as broad as that of high surface brightness galaxies. The latter will be referred to below as ``normal'', or HSB-galaxies. 

In this paper, we restrict our consideration to non-dwarf LSB-galaxies ($M_{B} \lesssim -18^{m}$). In most cases, they belong to late-type spiral galaxies in which, despite their low surface brightness, young stellar populations are observed. As a rule, such galaxies contain large amounts of $HI$ for their luminosity, with very low lelative mass of  molecular gas. At the same time, while having roughly the same gas surface density as HSB-galaxies, low surface brightness galaxies are significantly inferior to the latter in star formation rates -- both integrated and per unit disk area (\cite{Lei_etal_2019}).

The reason for the low brightness and low star formation rate in LSB-galaxies is the subject of ongoing scientific debate. Several scenarios for the formation of their observed features have been proposed by different authors (see, e.g., the discussion in \citet{Bothun_1997, Das_2013}, \citet{Saburova_etal_2021, Saburova_etal_2023}, \citet{Du_etal_2023}). One of several possible scenarios is the external gravitational interaction and/or merging of initially HSB-galaxy with another galaxy, which led to the redistribution of angular momentum and the formation of  extended low-brightness disk with high specific angular momentum and low matter density. 

Another possibility is that LSB disk is not a result of  angular momentum redistribution, but is the sequence of ``growth'' of the outer disk in the already formed normal galaxy (e.g., NGC 289; see \citet{Walsh_1997}). In this case the appearance of the outer low-brightness disk is resulted either from the accretion of high-angular-momentum gas or from the disruption of dwarf satellites (minor merging) whose orbits were in a plane close to the disk plane.  

Finally, LSB disks may reflect the conditions of their formation in a protogalaxy with an extended dark halo of high specific angular momentum. In this case, one would expect that the equilibrium  disk which forms within the halo in a single process  will have a low radial density gradient and  low surface and volume densities. This scenario most likely applies to galaxies where the dark halo component of the rotation curve rises slowly with radius, showing a low central density concentration (\citet{Saburova_etal_2021}). In the scenarios described above, the low density of galactic disks leads to the extremely low star formation rates, and, as the result, a significant fraction of the gas in the disks may still remain unprocessed.

In HSB-galaxies, the distant regions of the disks are in general similar to LSB disks. In both cases, the factors stimulating star formation should be greatly weakened in a tenuous gas in a disk plane  (see \citet{Elmegreen_Parravano_1994}). Nevertheless, UV observations (GALEX) have shown that more than 10$\%$ of HSB-galaxies possess extended low-brightness outer disks containing young stellar populations that extend far beyond the conventionally defined optical radius $R_{25}$ (\citet{Thilker_etal_2007}, \citet{GilDePaz_2017}, \citet{Elmegreen_Hunter_2017}, \citet{Outskirts_proc}). In both LSB-galaxies and the outer disks of HSB-galaxies, a faint spiral structure may be observed, along with the individual small $HII$ regions at large distances from the center (see, e.g., \citet{Werk_etal_2010}, \citet{Barnes_etal_2012}). 

In the rarefied disks a relative mass fraction of molecular gas is expected to be extremely low, and the star formation rate is primarily linked to atomic gas. The most significant amount of molecular gas was detected in the disk of the giant LSB-galaxy Malin 2, but even there it’s mass does not exceed a few percent of the $HI$ mass (\citet{Das_etal_2010}). At the same time, the radial profiles of the $HI$ surface density and star formation rate in low-brightness regions vary greatly from object to object, showing different formation and evolutionary histories of the observed disks of galaxies. 

The open question is whether star formation conditions in LSB-galaxies and in the tenuous regions of HSB disks are similar. Some authors considered the possibility that the low brightness of LSB disks is simply a consequence of an anomalously high $M_{*}/L$ ratio of the stellar population, related to peculiarities of the initial mass function (top-light IMF, \citet{Fuchs_2003}, \citet{Lee_etal_2004}).  However, modeling of the stellar spectra of LSB galaxies did not reveal large differences between galaxies of different surface brightness when comparing their positions on the ``$M_{*}/L$–color'' diagram (\citet{Du_etal_2020}). A UV study of star-forming complexes in the extended (XUV) disk of the relatively nearby galaxy NGC 2090 also did not find a deficit of massive stars (\citet{Yadav_etal_2026}). 

Nevertheless, the question of the dependence of the $IMF$ of newly formed stars on the gas density or disk surface density is not closed (see the discussion in \citet{Hoversten_Glazebrook_2008}, \citet{Meurer_etal_2009}, \citet{Lee_etal_2009}, \citet{Kauffmann_2021}, \citet{Kasparova_etal_2014}). This issue applies equally to the peripheral regions of HSB-galaxies and to LSB disks as a whole. 

A comparison of the outer regions of HSB- and LSB-galaxy disks is the main subject of this paper.

\section{Galaxy Sample}

For comparing LSB-galaxies with the outer disk regions of normal galaxies, we use as the latter the spiral galaxies from the THINGS survey (\citet{THINGS}), as well as normal-brightness galaxies characterized by a high relative $HI$ mass (HI-rich, or HIR-galaxies) from Lutz et al. (\citet{Lutz_1}, \citet{Lutz_2}, \citet{Lutz_3}). These papers consider late-type galaxies containing 2.5 or more times of $HI$ than it would be expected based on their optical characteristics. The sample of normal-brightness galaxies also includes the high-gas-content spiral HSB-galaxies presented by (\citet{Huang_HIR}) as galaxies with the extended $HI$ disks. From the above-mentioned samples, those galaxies were selected, for which the radial $HI$ surface density profiles $\Sigma_{HI}$ and brightness profiles beyond the photometric radius $R_{25}$ are available in the literature.

This sample of LSB-galaxies includes non-dwarf disk galaxies with known rotation curves and moderate disk inclination angles.

Table~\ref{tab:sample} contains the complete list of galaxies, the adopted distances and  optical luminosities. In all cases, except for Malin 1, brightness profiles from the LEGACY survey (\citet{LEGACY}) were used to estimate luminosities. For Malin 1, photometric estimates were taken from \citet{Boissier_etal_2016}. Parameters depending on the adopted distance have been scaled to a distance scale of $H_{0} = 70$ km/s/Mpc. The normal-brightness spiral galaxies (HIR and THINGS samples) have absolute magnitudes in the range $-21.9 < M_{B} < -19.4$, while LSB- galaxies show a slightly larger spread in this parameter ($-21.8 < M_{B} < -17.8$) due to three galaxies of lower luminosity. Both HSB- and LSB-galaxies have high rotation velocities ($V_{rot} > 110$ km/s), with the exception of two LSB-galaxies (F561-1 and UGC 5750) and the HSB-galaxy NGC 628 (THINGS), which rotate somewhat more slowly.

To estimate star formation rates for LSB-galaxies, we used FUV surface brightness $\mu_{FUV}$ from GALEX data, corrected for disk inclination (following \citet{Wyder_etal_2009}, equation 4 therein), where $SFR$ was calculated for a Salpeter $IMF$ (\citet{Salpeter_1955}). For normal spiral galaxies, data on the radial distribution of $SFR$ were taken from \citet{Leroy_etal_2008}, where authors found $SFR$ per unit disk area from GALEX data, the same as for the LSB-galaxies, but for a different initial mass function (Kroupa-type two-component IMF). To bring the estimates to the same $IMF$ as for the LSB-galaxies, a conversion factor of $\sim$2.1 was used (see  equation (10) in \citet{Salim_etal_2007}). When comparing $SFR$ for regions far from the center, HIR-galaxies (these are HI-rich HSB-galaxies from  Lutz et al.) were not included due to a lack of data on the radial distribution of $SFR$ for them.

Data sources on $HI$ surface density and galaxy rotation curves are given in Table~\ref{tab:sample}. Values of optical disk diameters ($D_{25}$), disk inclination angles ($incl$), and galaxy radial velocities ($V$) were taken from the HyperLeda database (\citet{Hyperleda}). For LSB-galaxies, a preference was given to the inclinations obtained from  kinematic analyses of the disks. For the THINGS galaxies, distance and inclination data were taken from \citet{deBlok_THINGS} (Table 2 in their paper).  

Interacting galaxies and galaxies with close companions of comparable luminosity were excluded from all subsamples to minimize the influence of tidal effects on star formation and gas distribution. In addition, we did not consider objects whose disks are viewed nearly face-on. These restrictions reduce the uncertainty in determining parameters related to the gas density or velocity of rotation radial profiles.

\newpage
\begin{longtable}{|l|c|c|c|c|c|}
\caption{Galaxies sample\label{tab:sample}} \\
\hline 
Name & $R_{25}$, kpc  & $R_{HI}$, kpc & $V_{rot}$, km/s & $incl$, deg & $M_B$ \\
\hline
\endfirsthead 
\hline 
\endhead 
\hline 
\multicolumn{2}{r}{\textit{Continued on the next page}} \\
\hline 
\endfoot
\hline 
\endlastfoot 
\hline
\multicolumn{6}{|c|}{LSB} \\
\hline
F561-1 & 8.5 & $10.8^1$ & $50.5^1$ & $0^1$ & -18.05 \\
F568-1 & 9.4 & $18.5^1$ & $115^2$ & $26^2$ & -18.34 \\
F568-3 & 10.2 & $---^1$ & $125^2$ & $40^2$ & -18.69 \\
F579-v1 & 7 & $15.7^2$ & $112.5^2$ & $26^2$ & -18.65 \\
Malin1 & 10 & $73.5^3$ & $214.1^3$ & $38^3$ & -20.22 \\
Malin2 & 30.8 & $61^3$ & $320^4$ & $38^4$ & -21.82 \\
UGC1382 & 12.4 & $61^4$ & $280^5$ & $46^5$ & -20.19 \\
UGC5750 & 7.7 & $16.5^5$ & $78.9^3$ & $64^2$ & -17.79 \\
UGC5999 & 7.5 & $15^5$ & $111^2$ & $30^1$ & -18.08 \\
UGC6614 & 16.6 & $57.5^3$ & $205.2^3$ & $35^4$ & -20.72 \\
\hline
\multicolumn{6}{|c|}{Normal} \\
\hline
NGC628 &	11.1 &	$19.8^6$ &	$64.3^6$ &	$19.8^1$ &	-20.68 \\
NGC2403 &	9.3 &	$15^6$ &	$120.1^6$ &	$61.3^1$ &	-19.4 \\
NGC2841 &	13.9 &	$34^6$ &	$319.6^6$ &	$65.3^1$ &	-21.53 \\
NGC2903 &	15.6 &	$21.3^6$ &	$189^6$ &	$67.1^1$ &	-21.02 \\
NGC3351 &	10.6 &	$12.3^6$ &	$150.7^6$ &	$54.6^1$ &	-19.84 \\
NGC3521 &	13.2 &	$23.7^6$ &	$245^6$ &	$60^1$ &	-21.25 \\
NGC3627 &	13.8 &	$8.2^6$ &	$174.7^6$ &	$67.5^1$ &	-21.21 \\
NGC4736 &	5.3 &	$4.2^6$ &	$181.6^6$ &	$31.8^1$ &	-19.67 \\
NGC5055 &	17.3 &	$20^6$ &	$218.1^6$ &	$54.9^1$ &	-20.89 \\
\hline
\multicolumn{6}{|c|}{Rich HI} \\
\hline
PGC624 & 	23.5 &	$48.9^7$ &	$197.3^7$ &	$57.9^1$ &	-21.34 \\
PGC2887 &	16.4 &	$56^7$ &	$166.5^7$ &	$39.9^1$ &	-20.93 \\
PGC3089 &	11 &	$77.5^7$ &	$167.8^7$ &	$43^1$ &	-20.2 \\
PGC7298 &	23.5 &	$51.1^7$ &	$189.9^7$ &	$79.7^1$ &	-20.99 \\
PGC11691 &	19.9 &	$45.6^7$ &	$173.9^7$ &	$50.7^1$ &	-20.6 \\
PGC33645 &	23.8 &	$36.6^8$ &	$193^8$ &	$80.1^1$ &	-21.33 \\
PGC42836 &	36.8 &	$38.2^8$ &	$221^8$ &	$80.1^1$ &	-21.83 \\
PGC50455 &	15.1 &	$39^9$ &	$174^8$ &	$60^1$ &	-20.92 \\
PGC70281 &	23.7 &	$59.6^7$ &	$179.1^7$ &	$90^1$ &	-20.87 \\
PGC71078 &	24.3 &	$38.5^9$ &	$230^8$	& $83.5^1$ &	-21.05 \\
\hline
\end{longtable}

\textbf{Notes:} Data sources for $HI$ distributions: 1 -- \citet{Abramova_2012}, 2 -- \citet{deBlok_etal_1996}, 3 -- \citet{Pickering_1997}, 4 -- \citet{Hagen_etal_2016}, 5 -- \citet{vanderHulst_etal_1993}, 6 -- \citet{Bigiel_etal_2010}, 7 -- \citet{Lutz_2}, 8 -- \citet{Hallenbeck_etal_2016}, 9 -- \citet{Hallenbeck_etal_2014}; kinematics and inclinations: 1 -- \citet{Haghi_etal_2018}, 2 -- \citet{Ghari_etal_2019}, 3 -- \citet{Capistrano_Barrocas_2018}, 4 -- \citet{Kasparova_etal_2014}, 5 --\citet{Hagen_etal_2016}, 6 -- \citet{deBlok_etal_2008}, 7 -- \citet{Lutz_2}, 8 -- \citet{Huang_HIR}; inclination data: 1 -- \citet{Hyperleda}, 2 -- \citet{deBlok_etal_2001}, 3 -- \citet{Lelli_etal_2010}, 4 -- \citet{Pickering_1997}, 5 --\citet{Hagen_etal_2016}; absolute magnitudes correspond to B-band magnitudes from HyperLeda \citet{Hyperleda}.
\clearpage

\section{Gas Density and Star Formation Rate in Low-Brightness Regions}

\begin{figure}[h] 
\begin{minipage}[h]{0.99\linewidth}
\center{\includegraphics[width=1\linewidth]{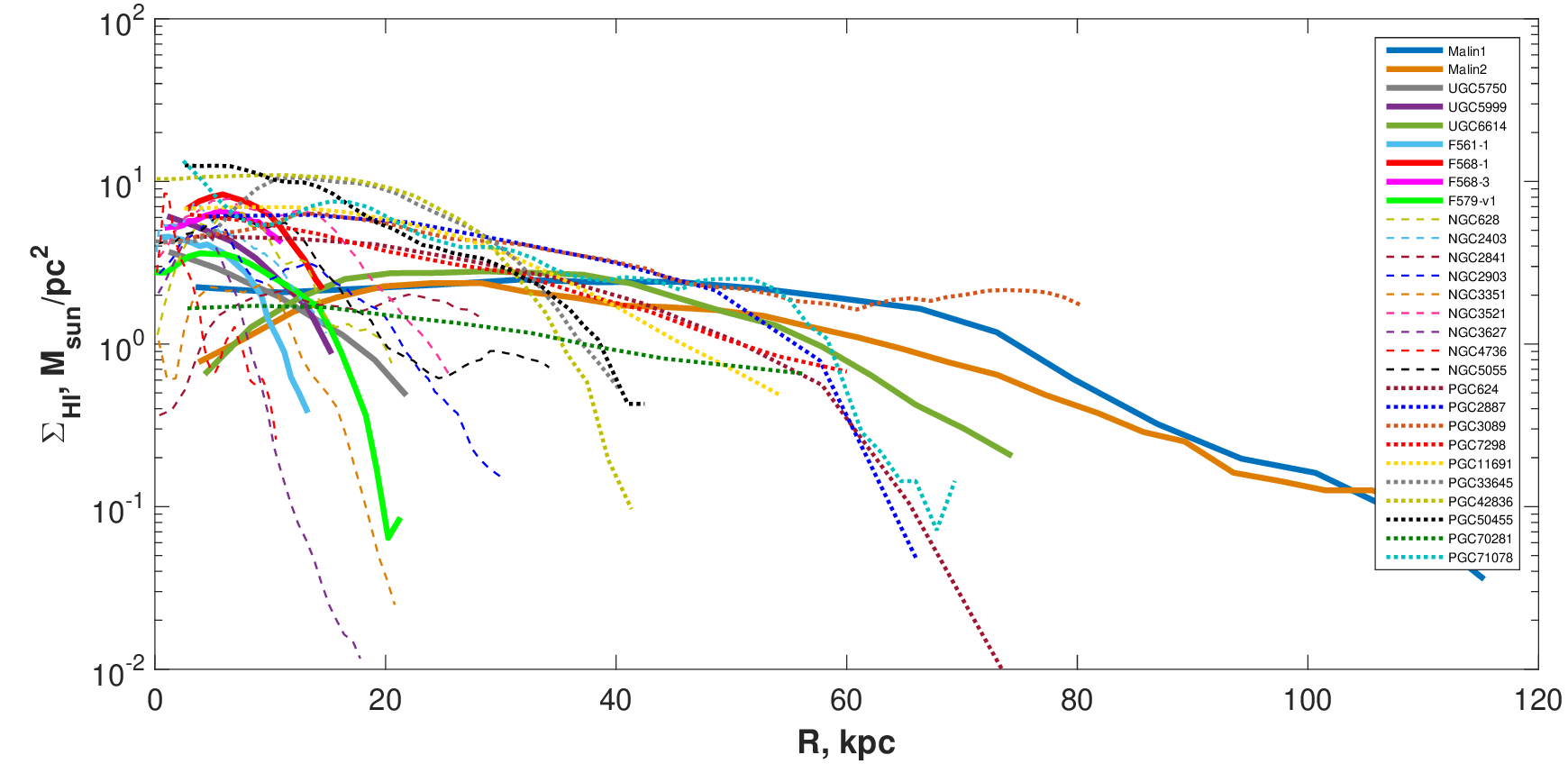}} a) \\
\end{minipage}
\hfill
\begin{minipage}[h]{0.99\linewidth}
\center{\includegraphics[width=1\linewidth]{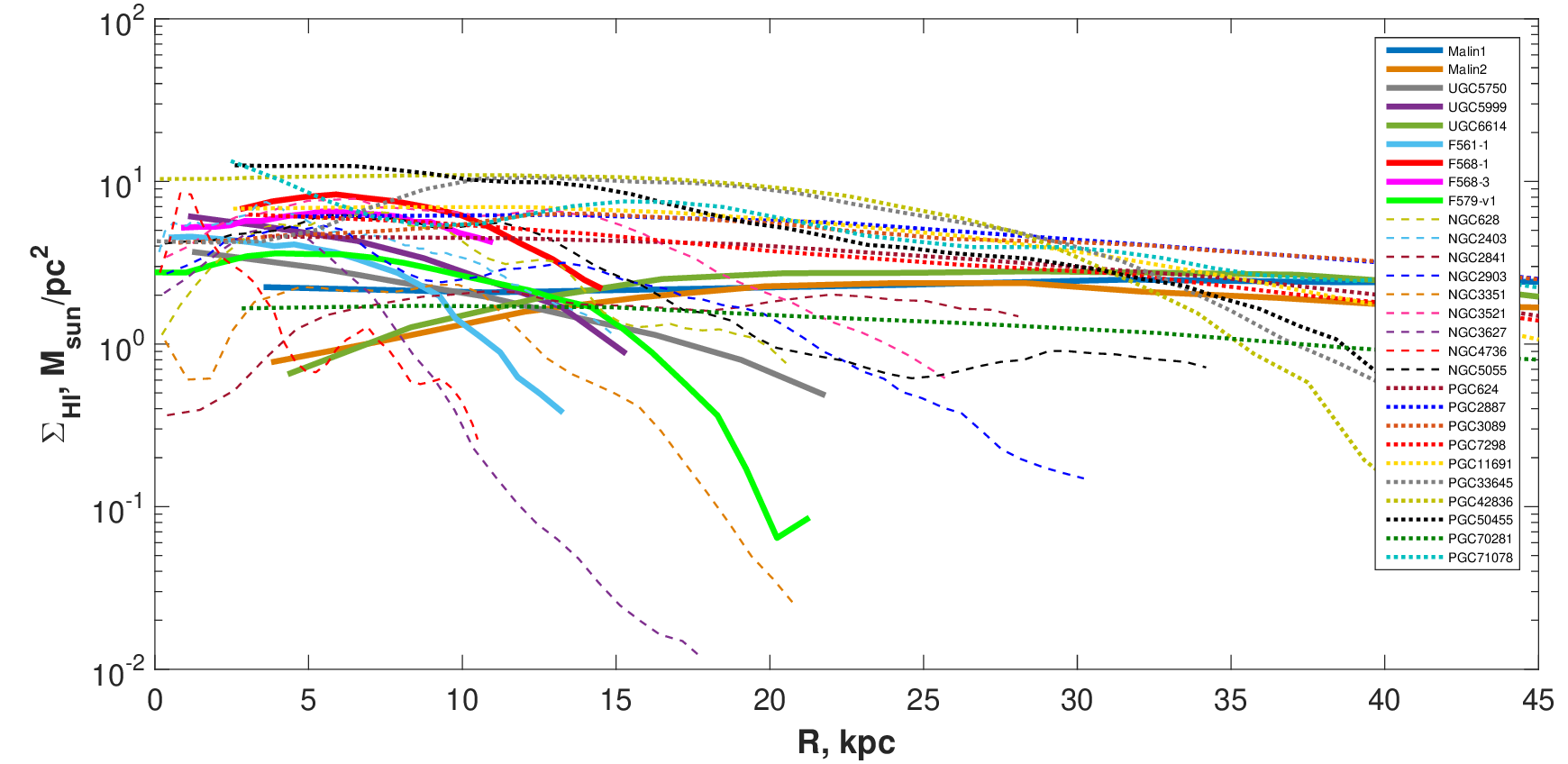}} b) \\
\end{minipage}

\caption{Radial profiles of $HI$ surface density for the galaxies considered. Solid lines are LSB-galaxies, dashed lines are THINGS galaxies, and short-dashed lines are HIR-galaxies. Panels (a) and (b) differ only in their scale of the horizontal axes.}
\label{fig:H_R}
\end{figure}

One can expect that the gas layer thickness increases along the radial distance due to the decreasing of density of the disk. For this reason, star formation in LSB disks and in the outer disks of HSB-galaxies occurs under conditions of very low average volume density of interstellar medium. Hence a local values of star formation rate per unit disk area ($\Sigma_{SFR}$) and star formation efficiency $SFE = \Sigma_{SFR}/\Sigma_{HI}$ there should be significantly lower than the mean values for a  galaxy as a whole.

A formation of gravitationally bound gas regions as the precursors of young stellar population  depends, besides a gas density, on a number of other factors. These include a presence of non-circular gas motions leading to gas stream collisions (e.g., in spiral arms), a metallicity of gas (which affects the cooling rate), and a radial gradient of  disk rotation velocity. Higher gas density and gas metallicity promotes star formation, while higher angular velocity makes the disk gravitationally stable counteracting a growth of condensations. It is essential that giant LSB-galaxies are characterized by lower angular rotation velocities in the outer disk regions compared to HSB-galaxies, which is a positive factor for star formation. Indeed, their disk rotation velocities are roughly similar to those of HSB-galaxies of the same  luminosity, while their disk sizes (a radial scale lengths) are significantly larger than in the latters (\citet{Du_etal_2019})).

A comparison of the observed star formation rates in HSB- and LSB-galaxies has been considered in a number of studies. \citet{Boissier_etal_2008} concluded that the star formation efficiency $SFE$ for LSB-galaxies is significantly lower than for HSB-galaxies. However, these authors used the integrated (total) values of $SFR$ and $M_{HI}$, whereas the radial distributions of these quantities differ greatly for  galaxies of different surface brightness. The dependencies of $SFR$ on the azimuthally averaged surface and volume densities of $HI$ of HSB and LSB-galaxies were  first compared in \citet{Abramova_Zasov_2012}. In this paper the volume densities of gas and stars were computed for the equilibrium  gravitationally stable disk (\citet{Abramova_Zasov_2011}). It was shown that $SFR$ correlates  with the average (at a given radius) volume density of gas in the disk plane, decreasing with distance from the center in a similar way for LSB and HSB-galaxies, although the relationship between these quantities is nonlinear. A universal character of the dependence between $SFR$ and the volume gas density in the disk plane over a wide range of disk densities was most clearly demonstrated in later works by  (\citet{Bacchini_etal_2020} and  \citet{Du_etal_2023}).

Regarding the star formation efficiency $SFE$, this parameter is more closely related not to a gas density but to a disk density. Moreover, as first noted in \citet{Abramova_Zasov_2011}, LSB-galaxies extend the relation established for HSB-galaxies into the low-density regime. The dependence of $SFE$ on a disk density is in a good agreement with the models of Ostriker et al. (\citet{Ostriker_Kim_2022} и \citet{Ostriker_etal_2010}) for self-regulated star formation in a quasi-equilibrium gas layer, where $SFR$ is linearly related to the gas pressure in the disk plane (see also \citet{Blitz_Rosolowsky_2006}).
  
Although a presence of a stellar disk, whose gravity compresses a gas layer, is the essential factor which determines $SFE$, this is neither a necessary nor a sufficient condition for star formation, since young star population in many cases can  be observed outside galactic disks. The examples are tidal tails of interacting galaxies,  or gas flows ejected from a disk due to ram pressure stripping (in jellyfish galaxies), containing $HII$-regions.

The most obvious factor enhancing the local process of star formation in galactic disks is gas compression associated with spiral arms. A faint spiral structure is often presented in low-brightness disks,  nevertheless a clearly defined spiral structure there  is more the exception than the rule, both for LSB- and HSB-galaxies. The appearance of local condensations in the gas layer can also be caused by tidal effects from neighboring galaxies, leading to non-circular gas motions, and, as a result, to gas stream collisions. However, the very existence of the extended gaseous disks with regular rotation already rules out strong tidal effects. This applies equally to low-brightness galaxies and to the extended disks of HSB-galaxies. In both cases, the average gas density at a given radius  remains the key parameter determining the star formation rate.

A comparison of the azimuthally averaged $HI$ surface density for HSB- and LSB-galaxies, $\Sigma_{HI}(R)$, is illustrated in Figs.~\ref{fig:H_R}(a,b), which differ only in the horizontal axis scale. Here, as in other figures, LSB-galaxies are shown as thick curves, HSB-galaxies from the THINGS sample -- as dashed lines, and HSB-galaxies with high neutral hydrogen content (HIR-galaxies) from the Lutz et al.  (\citet{Lutz_1}) and Huang et al. (\citet{Huang_HIR}) samples - as dotted lines. The $HI$ densities in the inner disk regions for almost all galaxies considered lie within $2-20 M_{\odot}/$pc$^{2}$ regardless of their surface brightness. Three giant LSB-galaxies (Malin 1, Malin 2, UGC 6614) stand out for their extremely extended gaseous disks, which can be traced out to several tens of kpc from the center. 

Some HSB-galaxies in Fig.~\ref{fig:H_R}(a) also have gaseous disks extending far beyond the optical radius $R_{25}$. As in LSB-galaxies, they may exhibit a low-contrast spiral structure, or (e.g., in NGC 4736) an outer low-brightness ring. At the same time,  $\Sigma_{HI}$ density in HSB-galaxies with the extended disks can remain above $1 M_{\odot}/$pc$^{2}$ even at $R$ > 25–30 kpc.

The example of a relatively nearby HSB-galaxy with the extended gaseous disk is the spiral galaxy NGC 2841. In \citet{Ianjamasimanana_etal_2018}, based on THINGS data, the radial $HI$ density profile in this galaxy was traced down to an extremely low threshold density of $HI \sim 10^{18}$ cm$^{-2} \approx 10^{-2} M_{\odot}/$pc$^{2}$ at a distance of about 80 kpc from the galaxy center -- much farther than it is shown in Fig.~\ref{fig:H_R}. However, no clear signs of star formation are observed at such low densities. 

Among the HSB-galaxies considered, the most extended gaseous disk, similar to those of giant LSB-galaxies, is found in PGC3089 (= NGC289), the SBbc galaxy with high $HI$ mass. The faint outer spiral arms of this galaxy extend out to 4 arcmin ($R \sim 20$ kpc), while the gas maintains a density of $\sim 2 M_{\odot}/$pc$^{2}$ out to $R \sim 80$ kpc. 

\begin{figure}[h]
\includegraphics[width=0.99\hsize]{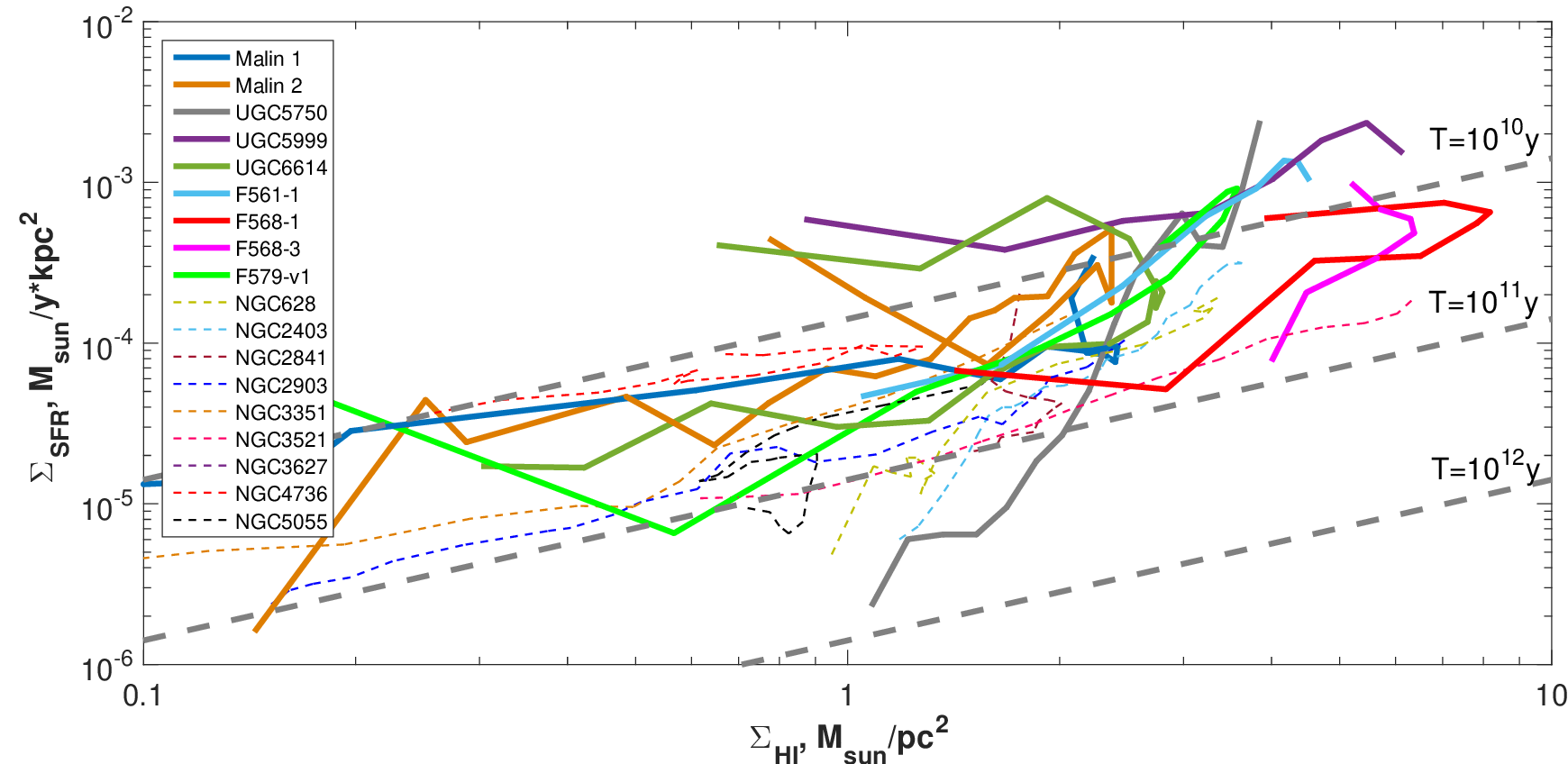}
\caption{\label{fig:SF_HI}
Dependence of star formation rate on $HI$ surface density for LSB-galaxies and the outer regions of THINGS galaxies. Labels are the same as in Fig.~\ref{fig:H_R}. The dashed lines correspond to different gas depletion times $T_{c}$.}
\end{figure}

In Fig.~\ref{fig:SF_HI} the star formation rates in LSB- and HSB-galaxies are compared with the gas surface density. The dashed lines correspond to fixed values of $SFE$, or equivalently, to the gas depletion time $T_{c} = \Sigma_{HI}/\Sigma_{SFR}$. It can be seen, that there is no threshold value of $HI$ density at which the star formation rate would drop sharply. 

The only exception is the LSB galaxy UGC 5750, where $SFR$ drops rapidly with decreasing $\Sigma_{HI}$. This is a late-type galaxy with the lowest luminosity among the galaxies considered which has a rather low rotation velocity (about 90 km/s). It differs from the other galaxies by an unusually narrow, high-contrast and extended bar, which indicates clearly atypical conditions for its structural formation.

Importantly, there is no systematic difference between the dependencies constructed for LSB- and HSB-galaxies , although the former show a larger scatter in $\Sigma_{SFR}$ values. 

\begin{figure}[h]
\includegraphics[width=0.99\hsize]{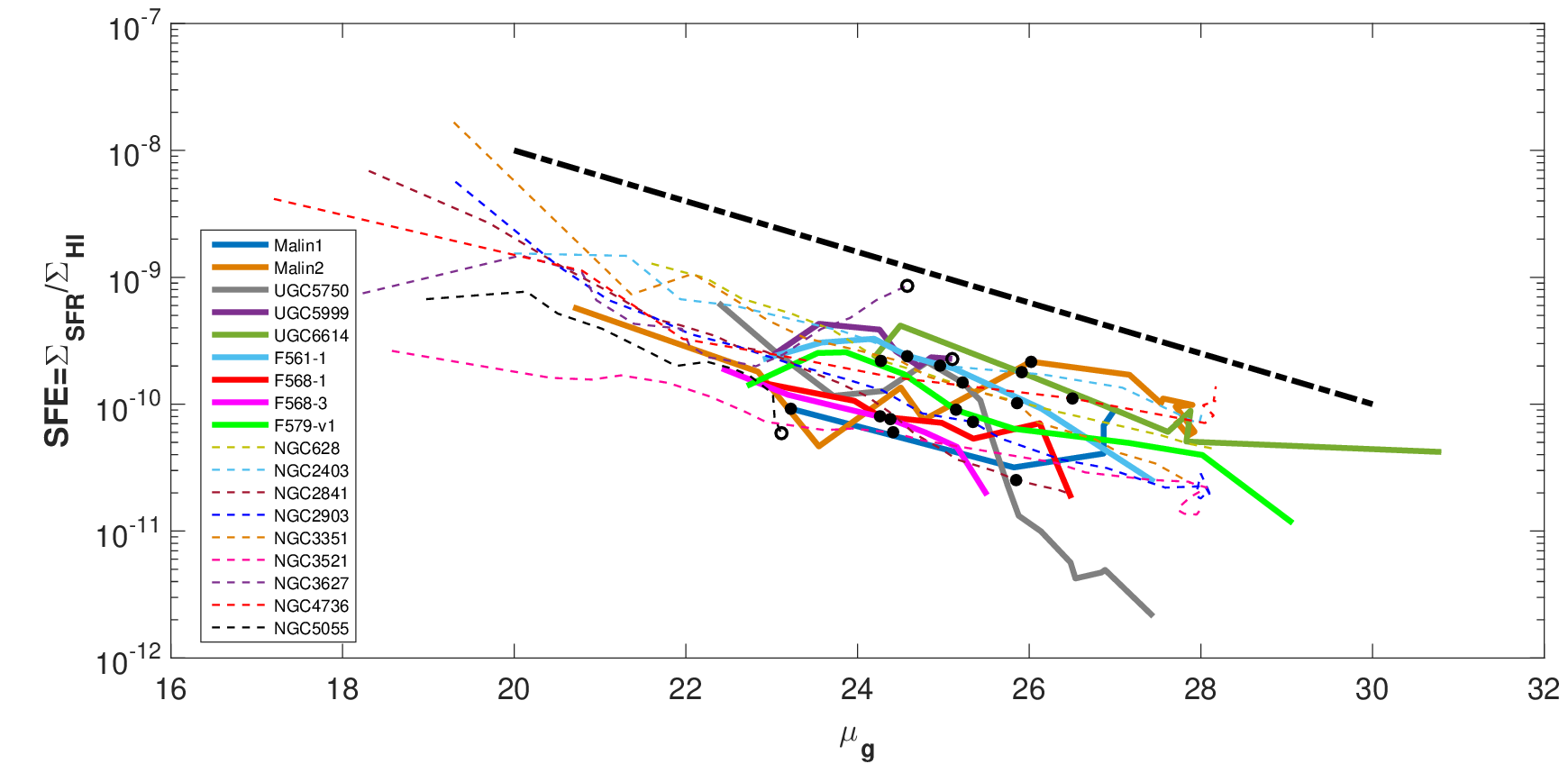}
\caption{\label{fig:SFE_g}
Star formation efficiency $SFE$ vs disk surface brightness $\mu_{g}$. Labels are the same as in Fig.~\ref{fig:H_R}. The slope of the dash-dotted line corresponds to the dependence $SFE \sim \Sigma_{*}^{1/2}$ for $M_{*}/L = const$. Black points on the curves mark the approximate surface brightness at which the surface densities of gaseous and the stellar disk become comparable. Open black circles mark galaxies where a  stellar disk surface density dominates over the entire range of considered radial distances.} 
\end{figure}

The variation of the star formation efficiency as a function of the optical disk brightness $\mu_{g}$ in the $g$ band for the galaxies considered is illustrated in Fig.~\ref{fig:SFE_g}. The brightness distribution approximately reflects the surface density distribution of a stellar disk. The systematic decrease of $SFE$ with $\mu_{g}$ is a natural consequence of the rapid drop in gas volume density with distance from the center. 

As it follows from the Figure, there exists a close relationship between $SFE$ and the surface brightness of  stellar disks. The relations in Fig.~\ref{fig:SFE_g} for different galaxies have approximately the same slope, with no systematic difference between LSB- and HSB-galaxies. The scatter in $\log SFE$ at a fixed surface brightness is $\sim$0.3 dex. A sharp decline in $SFE$ at the periphery of the optical disk is observed only for the LSB-galaxy UGC 5750, which also exhibits an anomalously low $SFR$ for its gas density (Fig.~\ref{fig:SF_HI}). 

If the $M/L_{g}$ ratio for galactic disks remains roughly constant along the radius, then the slope of the relations in Fig.~\ref{fig:SFE_g} corresponds to a simple power law $SFE \sim (\Sigma_{*})^{1/2}$ (parallel to the dash-dotted line).   
\

In LSB-galaxies, just as in the case of the extended outer disks of HSB-galaxies, a total surface density of disks $\Sigma_{d}(R) = \Sigma_{*}(R) + \Sigma_{gas}(R)$ decreases along the radius, while the ratio $\Sigma_{gas}/\Sigma_{*}$ increases, since the gas surface density decreases more slowly than the disk brightness. The black circles on the curves shown in Fig.~\ref{fig:SFE_g} mark the brightness $\mu_{g}$ at which the surface densities of gas and the stellar disk are equal for the accepted $M/L_{g}$ ratio of 0.5 solar units. This ratio corresponds to the average value for stellar disks found from the spectral energy distribution (SED) fitting for low-brightness galaxies (\citet{Du_etal_2020}). Open circles at the ends of the lines for several galaxies indicate that over the entire considered $\mu_{g}$ range, the stellar disk surface density remains higher than that of the gaseous one.  

Assuming that the scale height of a stellar disk $h_{*}$ is not smaller than the height of the $HI$ layer, $h_{g}$, one can estimate a lower limit for the latter. For the line-of-sight velocity dispersion of gas clouds $\sigma_{z} \sim 7 –10$ km/s, corresponding to a temperature of the intercloud  medium of about $10^4$ K, the formally calculated scale height of gas layer $h_{g} = \sigma_{z}^{2}/\pi G\Sigma_{d}$ with a surface density $\Sigma_{d} < 2 M_{\odot}/$pc$^{2}$ would be several kpc, and the volume gas density in a disk plane would be less than $10^{-26}$ g/cm$^3$. The occurrence of star-forming regions at such low gas density is problematic. 

Note that these order-of-magnitude estimates do not account for the role of the halo, whose gravitational field slightly ``squeezes'' the gas layer. The distance from the center at which the halo's role becomes significant or dominant can be estimated from a simple consideration. In the absence of a halo, the density of a self-gravitating equilibrium disk $\rho(z)$ decreases with $z$ as $\rho(z) = \rho_{0} sech^{2} (z/z_{0})$, where $z_{0}$ characterizes the disk scale height. The volume density of the disk, equal to $\rho_{0}$ in the disk plane, decreases slowly with $z$ (for $z \leq z_{0}$), and the acceleration along $z$ axis in the gravitational field of the disk can be taken approximately as $a_{d} \approx 4\pi G\rho_{0}z = 2\pi G\Sigma_{d}$. In the presence of a spherical halo which mass is   $M_{h}(R)$ within a sphere of radius $R$, a gravitational field of  a halo adds to the acceleration $a_{d}$ the quantity
\begin{equation}\label{ah}
a_h(z)= GM_h(R)/R^2\cdot (z/R) \approx \Omega_h(R)^2 z,
\end{equation}
where $\Omega_{h}(R) = V_{h}(R)/R$ is the angular velocity of the disk with a ``flat'' rotation curve $V_{h} \approx const$, which is close to the observed angular velocity of the outer disk $\Omega(R) = V_{rot}/R$. Importantly, both acceleration components - $a_{d}$ and $a_{h}$, associated with the disk and the halo respectively - are proportional to $z$. From these relations it follows, that these components become comparable when the disk surface density is
\begin{equation}\label{Eq2}
\Sigma_d(R) \approx \sigma_z (V_{rot}/R)/\sqrt{2} \pi G.
\end{equation}
The typical angular velocities of the peripheral regions of the disks considered here are (1–3) $\times 10^{-16}$ sm$^{-1}$. At these velocities, the halo's influence on the disk thickness becomes dominant starting from disk surface densities of several $M_{\odot}/$pc$^{2}$. For normal brightness galaxies, this corresponds in most cases to disk regions beyond the photometric radius $R_{25}$, while for LSB-galaxies it may apply to the entire disk except for the brightest inner region, usually occupied by a bulge. Therefore, the dark halo is a factor significantly limiting the thickness of the gas layer in the outer low-density disk regions.  

If a dark halo is not a sphere but rather a flattened ellipsoid, the halo's role becomes even more important. However even in the case of a strong flattening, the gas layer must expand outwards, so the gas volume density  decreases outward. Therefore, the star formation rate and the efficiency under low gas density is determined by the probability of formation (due to internal or external factors) of local regions of enhanced density in tenuous gaseous disks.

\section{How Far Are Galaxy Disks from the Gravitational Stability Threshold?}

The threshold density corresponding to the  marginal stability of the gas layer depends on the velocity dispersions of both the gaseous medium $c_{g}$ and the stellar disk $c_{*}$. The relatively small size of star-forming regions dominating in LSB disks and in the outskirts of normal spiral galaxies (\citet{Herbert-Fort_2012}) makes the gravitational instability of the disk as a mechanism for their formation unlikely -- at least at the current stage of galaxy evolution, since at the galaxy periphery, due to the low disk density, instability can develop only on large scales with a characteristic dimentions $\lambda_{crit} = 2c_{g}^{2}/G\Sigma_{disk}$ of several kiloparsecs (see also \citet{Barnes_etal_2012}).  
 
Nevertheless, due to the low star formation rate, the extended disks of LSB-galaxies can maintain the density level close to the threshold value for billions of years. This critical density value corresponds to the Toomre parameter $Q_{T,crit} = c_{g}\varkappa/\pi G\Sigma_{d}$, which is unity for axisymmetric perturbations of a thin one-component disk and is 1.5–2.5 for two-component disks (stars+gas) (\citet{Elmegreen_2011, Griv_Gedalin_2012, Marchuk_Sotnikova_2018}). Here $\varkappa$ is the epicyclic frequency, which for a ``flat'' rotation curve is $\sqrt{2}V_{rot}/R$. The finite thickness of disks makes them more stable, reducing $Q_{T,crit}$, while non-axisymmetric perturbations increase $Q_{T,crit}$ by a factor of 1.5–2.

A lack of reliable data of stellar velocity dispersions at large radial distances leads to a very approximate estimates of the Toomre parameter $Q_{T}$ for the low-density stellar disk. The situation, however, is simplified by the fact that in the presence of an extended gaseous disk a contribution of a stellar component to a general gravitational field  decreases toward the periphery, so at large radial distances  a stellar component  may not play a significant role in a disk stability condition.

The observed $HI$ velocity dispersion decreases slowly with $R$, and for the outer disk regions, in both HSB- and LSB-galaxies, it is typically 7–10 km/s (see, e.g., \citet{Pickering_1997, Meurer_etal_1996, Elson_etal_2010, Tamburro_etal_2009, Yadav_etal_2021}). The velocity dispersion of stellar component  $c_{*}$ is higher than that of the gas $c_{g}$, at the same time it decreases toward the periphery. To determine the critical value of the Toomre parameter $Q_{T}$, we adopt a fixed value of $c_{g} = 10$ km/s for the gas component, and for the lower limit of stellar velocity dispersion, so the disk as a whole can be considered as a single-component. In this case, the following inequality holds:
\begin{equation}\label{Eq3}
Q_{T} \ge \frac {c_g \varkappa}{\pi G (\Sigma_{gas}+\Sigma_*)}.
\end{equation}

The epicyclic frequency $\varkappa$ entering (\ref{Eq3}), is proportional to the angular rotation velocity of the disk $V/R$ (for a ``flat'' rotation curve, $\varkappa(R) = \sqrt{2}V/R$).

To take into account a stellar component of a disk, we used the photometric profiles of  galaxies, adopting the $M/L_{g}$ ratio of 1 (in solar units) as an upper limit (see, e.g., \citet{Into_Portinari_2013}). 
Estimates of the mass-to-light ratios for LSB disks obtained from spectral modeling confirm that for the vast majority of low-brightness galaxies, the $M_{*}/L$ values lie within the range 0.3–0.8 solar units, averaging $\sim$0.4–0.5, and slightly lower for dwarf LSB-galaxies; moreover, no clear differences are observed between these ratios for LSB- and HSB-galaxies (\citet{Du_etal_2020}). Since in most of the galaxies considered, the gas component density is comparable or exceeds the stellar density in the outer disk regions, the value of $Q_{T}$ for rarefied  galaxy disks depends weakly on the adopted $M_{*}/L_{g}$ ratio for both LSB- and HSB-galaxies.

\begin{figure}[h]
\includegraphics[width=0.99\hsize]{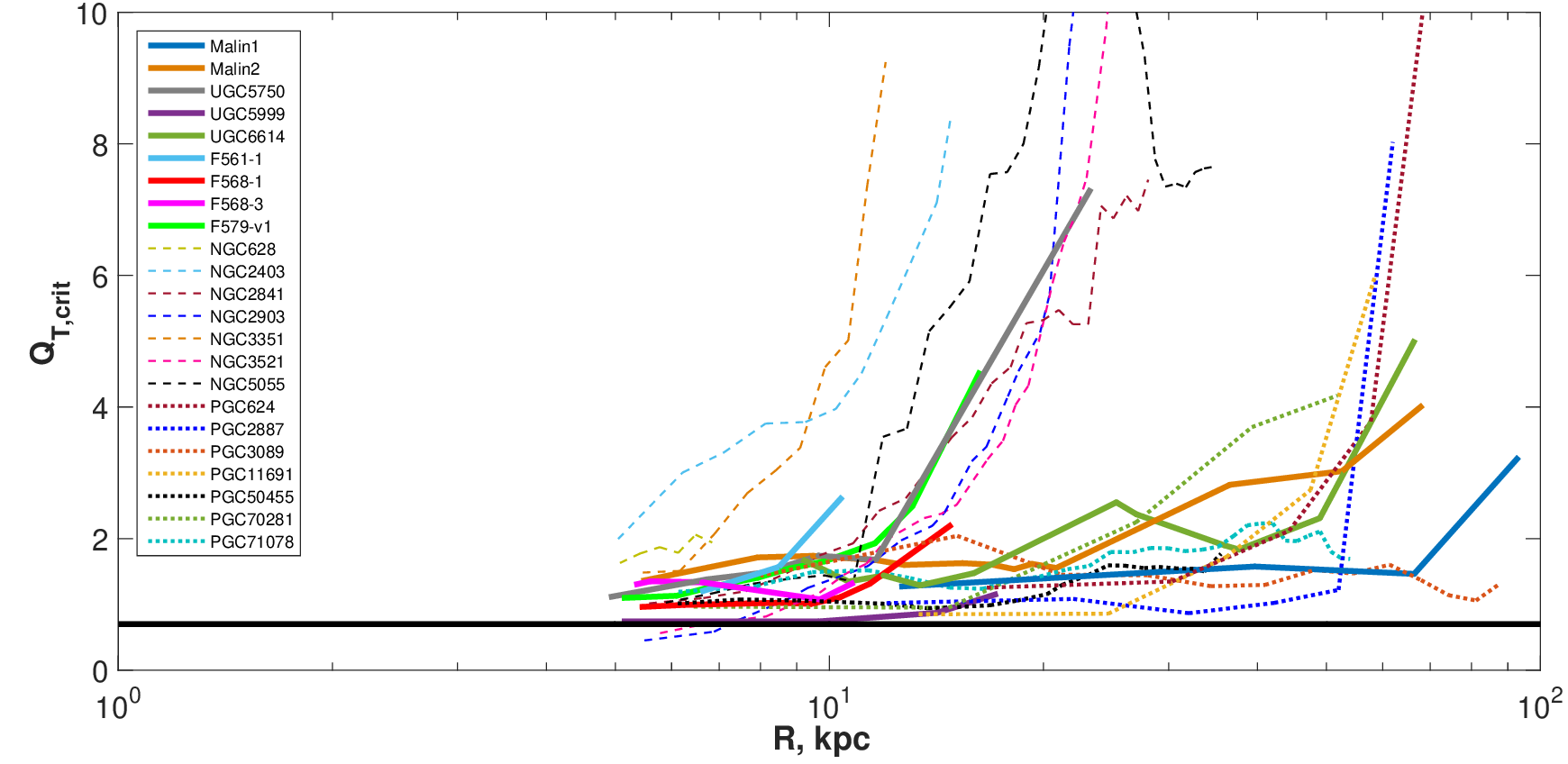}
\caption{\label{fig:Q}
 The radial variation of the Toomre stability parameter $Q_{T,crit}$ for the gas–stellar disk (see the text). Labels are the same as in Fig.~\ref{fig:H_R}.}.
\end{figure}

Figure~\ref{fig:Q} shows the radial variation of $Q_{T,crit}(R)$ for galaxies we consider with known radial profiles of stellar brightness and gas surface density. The horizontal line $Q_{T,crit} = 0.7$ is taken as the lower bound of $Q_{T,crit}$. The separation of the curves in the figure into two groups according to the radial extent of the disks is an effect of the inhomogeneity of the sample, which includes, along with the relatively small galaxies, the galaxies with the most extended gaseous disks.  

The radial distribution of $Q_{T,crit}(R)$ confirms that there is no evident difference between the outer regions of the extended gaseous disks  of HSB-galaxiesand the disks of  LSB-galaxies. In both cases, over a large range of $R$ (up to several tens of kpc!) the lower estimate of the stability parameter lies between 1 and 3, meaning that the disks are most likely close to the marginal stability despite their low surface density, without indication of a strong dynamical heating at a large range of $R$. The approximate constancy of $Q_{T,crit}$ for those galaxies where this quantity can be traced over a significant range of radii is a consequence of a reducing of angular velocity along the radial distance, which compensates by the decrease in disk density. At the far periphery of galaxies, due to the extremely low gas density, the Toomre parameter rise sharply (if the gas velocity dispersion does not fall significantly).

The existence of the extended low-brightness regions where the disk stability parameter varies little with $R$ suggests that during the formation of both LSB-galaxies and the outer regions of normal galaxy disks, gravitational instability could have played a determining role in establishing the observed radial density profile of the gaseous-stellar disk. After the initial phase of active star formation came to an end, a gas velocity dispersion slowed down to the observed speed of 7 - 10 km/c, and the disks settled into a quasi-stationary gravitationally stable state. Their star formation rate dropped to the present low values, corresponding to the gas depletion time $T > 10^{10}$ years (Fig.~\ref{fig:SF_HI}). As a result, the disk densities over a large range of $R$ remained  close to the critical values for stability for a long time. 

\section{Discussion and Conclusions}

The observed disks of both normal and low surface brightness galaxies may have different formation and evolutionary histories, so one can expect them to be very different in their rates and efficiencies of star formation. Nevertheless, as follows from Fig.~\ref{fig:SFE_g}, the current  $SFE$ remains closely related to the density of the stellar disk (see also \citet{Du_etal_2023}). At the same time, there is no systematic difference between normal and LSB-galaxies on the diagram: $SFE$ decreases with decreasing brightness approximately as $SFE \sim \Sigma_{*}^{1/2}$ (parallel to the dash-dotted line in Fig.~\ref{fig:SFE_g}). Such a dependence is consistent with the expected relation in the model of thermal and dynamical equilibrium of the interstellar medium embedded in a stellar disk of fixed thickness (\citet{Ostriker_etal_2010}; see also \citet{Shi_etal_2011}). Remarkably, it also holds in regions where the stellar disk density is lower than the gas density, where the role of  stellar component would seem to be diminished. This allows to  suggest that the surface density of low-brightness disks is probably underestimated.

Radial profiles of the azimuthally averaged star formation rate in the disks of galaxies shown in Fig.~~\ref{fig:SFE_g} are traced out to surface brightnesses of $\mu_{g}\approx \, $26–27$^m$/arcsec$^2$, corresponding to stellar disk surface densities of $\sim 0.5-2.5 M_{\odot}/$pc$^2$ for $M/L_{g} \sim 0.5-1$. The stellar disk density decreases along the radius much faster than the observed velocity dispersion of gas: at the disk periphery, the latter remains close to that expected for thermal velocities of a single-phase $HI$  with $T \approx 7000$ K. This should inevitably lead to a steep increase in the vertical scale height of the gas layer at the disk periphery -- from several hundred parsecs to a kiloparsec or more in some galaxies (see the discussion in Section III)  -- and, consequently, to a sharp decrease in gas volume density at the periphery. As a result, one would expect a steep decline in the star formation rate and efficiency in the low-density disk regions where the contribution of the stellar disk to the vertical potential gradient becomes small. 
Below we consider  two possible scenarios accounting for the absence of the expected steepening of the drop of  $SFE$ at the disk periphery in most of the galaxies.

\textbf{Scenario 1.} Star formation in the  extremely tenuous disk regions is weakly related to the average disk density at a given $R$, and therefore depends weakly on the azimuthally averaged disk density. In this scenario, star-forming regions in a tenuous disk arise from random local condensations in the gas layer. They can be associated both with gas accretion and minor satellite accretion, and with the existence of dark matter subhalos, which, passing through the disk, gravitationally perturb the gas layer. This is a perfectly plausible scenario, but it  can only apply to the most distant regions of the disk, where the star formation rate depends weakly on the azimuthally averaged gas density. 

\textbf{Scenario 2.} Contrary to expectations, a thickness of the extended gaseous disks does not experience the sharp increase at the periphery that would be for a low-density gas–stellar disk. Since the gas ($HI$) velocity dispersion in the outer disk regions remains fairly high, such a scenario points to a possible  underestimation of the density of a low brightness parts of galactic disks, which would be the case if the disk contains an additional component beyond the observed gaseous and stellar ones. Below we consider this scenario in more detail.

The idea of the underestimation of density of low-brightness disks in LSB-galaxies or in the outer regions of normal galaxy disks has been put forward by several authors to explain the observed spiral structure in tenuous disks (\citet{Fuchs_2003}), or to consider a possible marginal stability of LSB disks (\citet{ Saburova_2012}). Some authors have suggested that some fraction of mass of galactic disks belongs to the unobserved cold gas (see \citet{Kasparova_etal_2014}, \citet{ Revaz_etal_2009} and references therein). It's also worth mentioning the paper by  \citet{Das_etal_2020}, where the dynamical mass of the disks of seven spiral and irregular galaxies observed nearly face-on was estimated from the measured $HI$ velocity dispersions. These authors found the excess of dynamical disk mass over baryonic mass (by a factor of several)  for several small Im galaxies where the gas mass dominates over stellar mass. For the outer regions of the extended spiral galaxy disks, their result remained inconclusive.
  
The presence of a ``dark'' disk component in addition to the baryonic (gas–stellar) disk would imply a smaller thickness and, accordingly, a higher volume density of the equilibrium gas layer in the galactic plane. A decrease in disk thickness makes it less stable against gravitational perturbations. This facilitates the formation of structural features related to disk self-gravity, such as the outer rings or low-contrast gas spiral arms, which in some galaxies can be traced far beyond the optical radius. A striking example is the existence of the extended gaseous spirals and the traces of weak star formation around the blue compact galaxy NGC 2915 which is surrounded by a giant hydrogen disk with regular rotation and the gas velocity dispersion of about 10 km/s (\citet{Bruzzese_etal_2015}).
   
A directly unobserved disk component (a baryonic or non-baryonic) may constitute a significant, if not the major, part of the disk mass where the observed total gaseous and stellar density does not exceed a few $M_{\odot}/$pc$^2$. In LSB-galaxies, this surface density is generally satisfied over almost the entire disk, while in HSB-galaxies it applies only to their outer low-brightness regions. The role of the ``dark'' component could be played, for example, by a disk (not necessarily thin) of non-baryonic matter brought into the outer regions of the galaxy by the captured and disrupted satellites, or the dark matter was originally included in the disk since its formation time. The latter scenario was proposed in \citet{Fan_etal_2013}, where the authors proposed that a certain fraction of the ``dark'' matter  may have a property of a collisional gas (a partially interacting DM).

 \citet{Yasui_etal_2024} proposed that at the extremely low gas metallicities, the formation of the excess number of low-mass stars may take place due to a lower $IMF$ mass cutoff, which would lead to a higher estimate of the disk density. However, there are currently no observational grounds for such a scenario.

In any case, as it follows from the present work, regardless of whether the ``missing'' mass is present in tenuous disks, low-brightness galaxies and the outer regions of HSB disks are indistinguishable in terms of the efficiency of the observed star formation. Consequently, the conditions determining the intensity of  star formation in these regions must be similar.

\section{Acknowledgments}

The study was conducted under the state assignment of Lomonosov Moscow State University.

\section{Bibliography}
\bibliographystyle{maik}
\bibliography{example.bib} 

\label{lastpage}
\end{document}